\documentclass{PoS}

\title{Local chiral fermions }

\ShortTitle{Local chiral fermions }

\author{\speaker{Michael Creutz}%
         \thanks{This manuscript has been
authored under contract number DE-AC02-98CH10886 with the
U.S.~Department of Energy.  Accordingly, the U.S. Government retains a
non-exclusive, royalty-free license to publish or reproduce the
published form of this contribution, or allow others to do so, for
U.S.~Government purposes.}\\
        Brookhaven National Laboratory\\
        E-mail: \email{creutz@bnl.gov}}

\abstract{
Borici's construction of minimally doubled chiral fermions builds on a
linear combination of two unitarily related naive fermion actions.
Being strictly local, extremely efficient numerical implementation
should be possible.  The resulting system is symmetric under the
subgroup of the hypercubic group that preserves a major hypercube
diagonal.  The symmetry includes both parity even and odd
transformations, but allows for an anisotropy to appear at finite
lattice spacing.}

\FullConference{The XXVI International Symposium on Lattice Field Theory \\
        July 14 - 19, 2008\\
        Williamsburg, Virginia, USA}

\begin{document}

\section{Introduction}

I describe a strictly local fermion Dirac operator ${\cal D}(A)$ with
an exact chiral symmetry manifested in anti-commutation with
$\gamma_5$, i.e. $\gamma_5{\cal D}=-{\cal D}\gamma_5$.  By strictly
local I mean that only nearest neighbor terms appear in the fermion
action.  The operator describes two fermion flavors, the minimum
required for chiral symmetry to exist.  I develop this action as a
linear combination of two ``naive'' fermion actions, following a line
of reasoning similar to that presented by Borici \cite{Borici:2007kz}.

The theory is not symmetric under the full hyper-cubic group, but the
subgroup thereof that preserves one fixed direction up to a sign.
These symmetries include transformations of both even and odd parity.
On renormalization, interactions can introduce a lattice anisotropy at
finite cutoff.  

\section{Doubling and chiral symmetry}

Spontaneously broken chiral symmetry is fundamental to our
understanding of low energy hadronic physics.  Pions are elegantly
described as quantum mechanical waves propagating through a background
quark condensate.  In addition, chiral symmetry provides powerful
tools enabling extrapolations to the physical quark masses from the
heavier values currently practical in lattice gauge simulations.

These issues are deeply entangled with quantum mechanical anomalies
that eliminate one symmetry of the classical Lagrangian.  With $N_f$
flavors of quark, the naive axial $U(N_f)$ is reduced to $SU(N_f)$.
With only one flavor, all chiral symmetry is removed; so, multiple
flavors are necessary for chiral symmetry to be relevant.  Nielsen and
Ninomiya \cite{Nielsen:1980rz} have given a formal topological
argument that any lattice action with chiral symmetry must describe at
least two flavors.

If one ignores the anomaly and writes a simple lattice action that is
chirally symmetric, something must go wrong.  The usual result is that
the fermion field describes multiple species, and the extra species
cancel the anomalies.  The most naive discretization, which will play
an essential role below, involves 16 species in four dimensions.
Staggered fermions divide out an exact $SU(4)$ symmetry of the naive
formulation to reduce the multiplicity to four \cite{staggered}.  The
Wilson fermion \cite{Wilson:1975id} approach successfully removes all
doublers at the expense of breaking all chiral symmetries.  Elegant
newer approaches based on perfect actions \cite{perfect},
domain walls in five dimensions \cite{Kaplan:1992bt}, or the overlap
operator \cite{Neuberger:1997fp} do maintain much of the desired
chiral symmetry with arbitrary $N_f$, but involve computationally
expensive non-local actions.  Also, these approaches tend to obscure
the anomaly; for example, with the overlap one introduces a gauge
field dependent matrix $\hat \gamma_5$, the trace of which gives the
winding number of a given gauge configuration.

The Nielsen-Ninomiya theorem requires any chirally symmetric lattice
action to describe at least two species.  Actions which satisfy the
minimal doubling of just two have been known for some time
\cite{Karsten:1981gd,Wilczek:1987kw}, and have recently stimulated new
interest \cite{Borici:2007kz,Creutz:2007af,maryland,Cichy:2008gk}.
There are a variety of compelling reasons to revisit these actions.
First is the failure of the rooting procedure popularly used to reduce
the doubling of staggered fermions \cite{rooting}.  Second is the lack
of an exact chiral symmetry for Wilson fermions, complicating
extrapolations to physical fermion masses.  And third is the severe
computational demands of the domain-wall and overlap approaches.

Here I construct a minimally doubled fermion action closely following
Borici's formalism \cite{Borici:2007kz}.  I use a linear combination
of two unitarily equivalent naive fermion actions.  This combination
will be crafted so that only two of the original 16 doublers for each
action survive.

\section{Naive fermions}

The so called ``naive fermion'' approach plays a crucial role in the
following; so, I begin by reviewing this action.  Work with a
conventional hyper-cubic lattice with gauge fields implemented as
group elements on the links.  The details of the gauge part of the
action play no role here.  When a fermion hops in a forward direction
$\mu$ between neighboring sites, it picks up a factor of $\gamma_\mu
U$.  Here $\gamma_\mu$ is the usual Hermitean Dirac matrix while $U$
represents the gauge field on the corresponding link.  For the
reverse, or backward, hop on the same link, the contribution is
$-\gamma_\mu U^\dagger$.  The minus sign makes the fermion operator an
anti-Hermitean matrix when viewed as a matrix in the direct product of
the site space with the internal symmetry and spinor spaces.  I work
here with the convention of a unit hopping parameter; for the massless
case any other hopping parameter can be scaled away with a
redefinition of the fields.  I also work in lattice units so that all
site coordinates are integers and the lattice spacing does not appear
explicitly.

\begin{figure*}
\centering{
\includegraphics[width=2.5in]{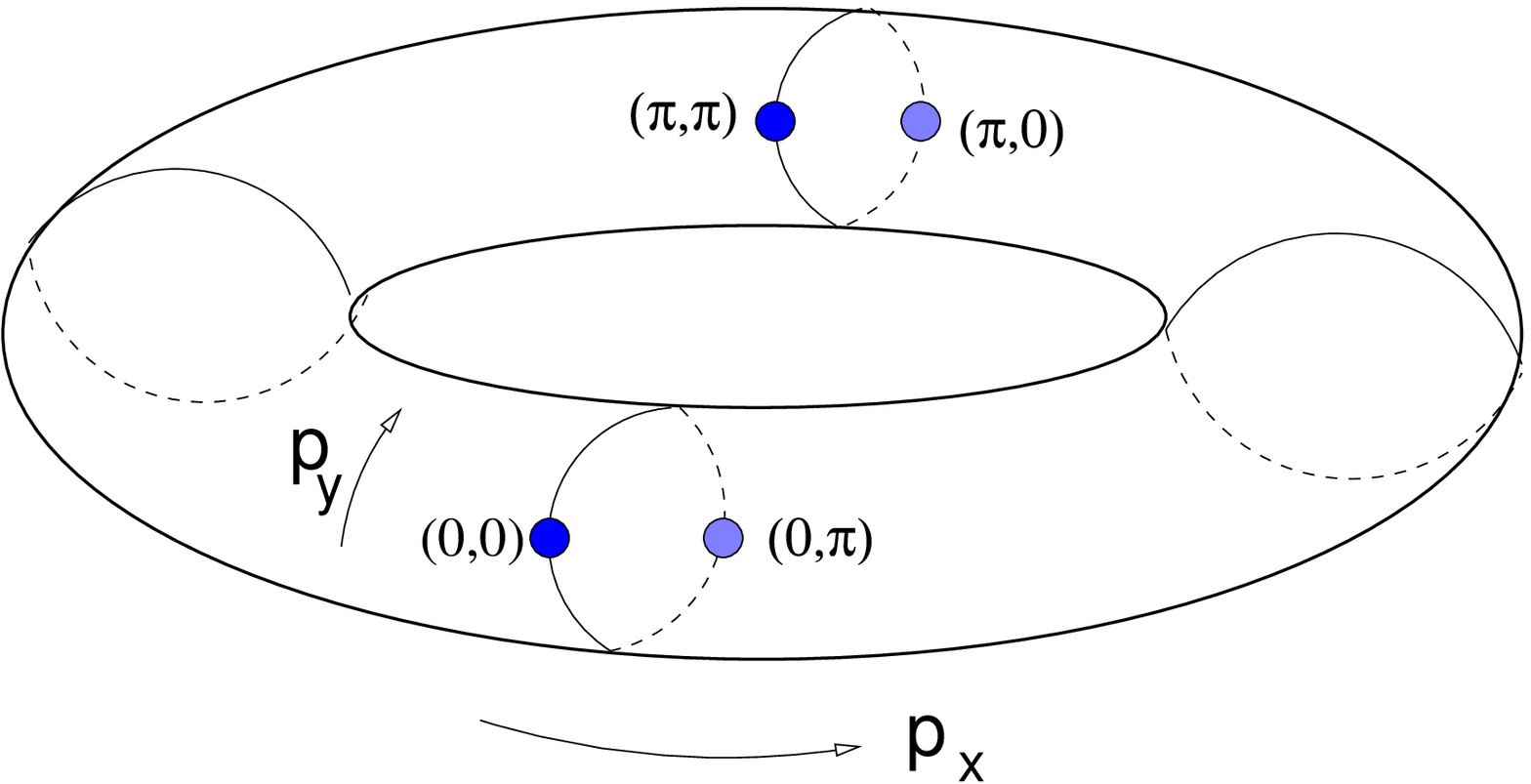}
\vbox to .07\vsize{
\hsize=.08\hsize
\centerline{$\times$} \vfill} 
\includegraphics[width=2.5in]{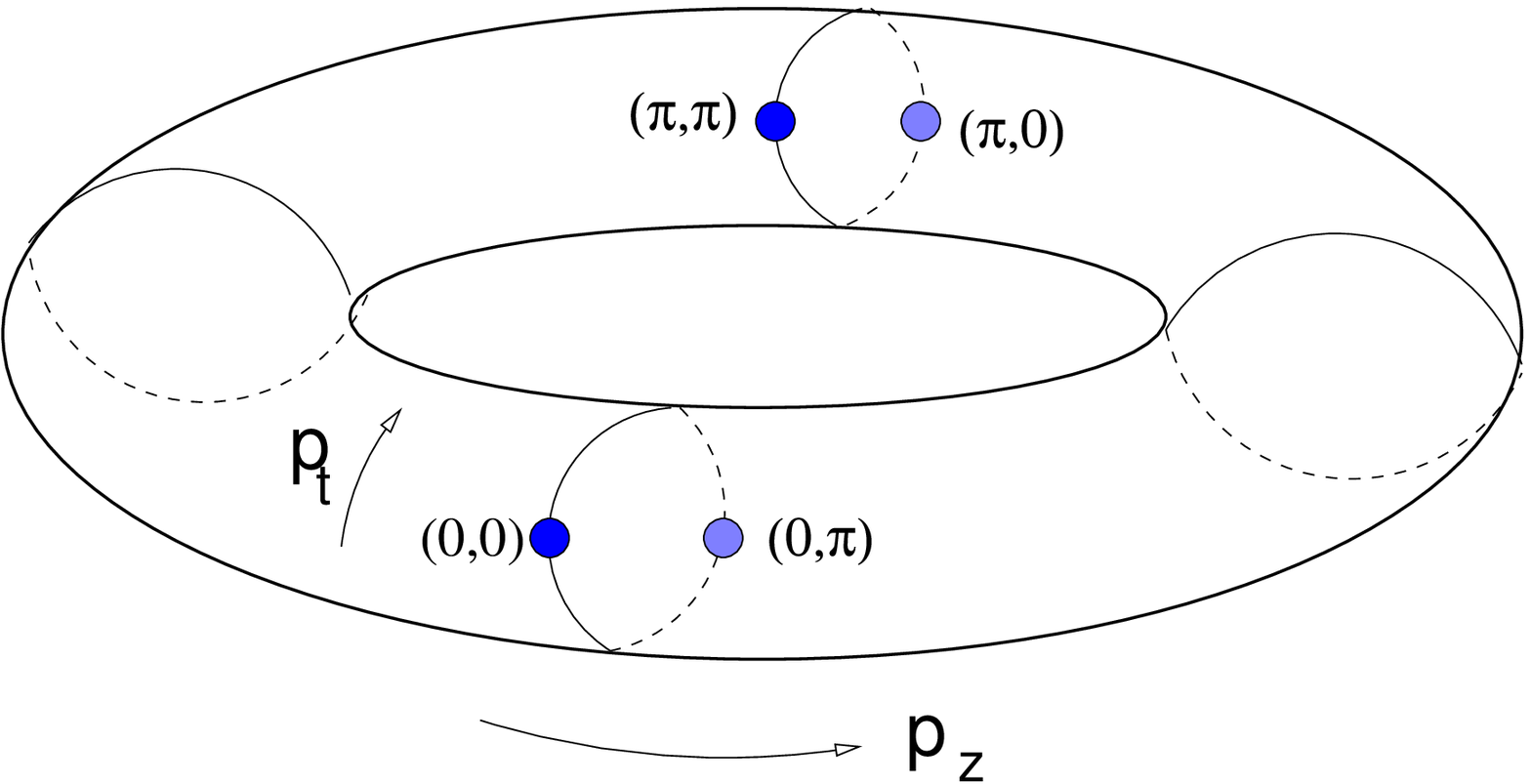}
}
\caption{Representing momentum space as a product of toroids, a
  massless Dirac equation can be obtained from naive fermions by
  expanding about any point where the components of momentum are
  either 0 or $\pi$.  This gives rise to the 16 doublers of the
  approach.}
\label{fig1} 
\end{figure*}

As is well known, this action describes the physics of 16 fermion
species, frequently referred to as doublers.  The Dirac operator, $D$,
anti-commutes with $\gamma_5$.  This represents an exact chiral
symmetry.  Because the doublers use different effective gamma
matrices, half of them rotate in each direction under a chiral
rotation.  Thus this is a non-singlet chiral symmetry.

In the free field limit where all the link matrices are the identity,
this theory is easily solved in momentum space.  The Dirac operator
factors into independent pieces for each momentum $p$, taking the form
\begin{equation}
D(p)=2i \sum_\mu \gamma_\mu \sin(p_\mu).
\end{equation}
Expanding about small momentum gives the usual Dirac behavior
$D(p)\simeq 2i \sum_\mu \gamma_\mu p_\mu$. The doublers appear on
expanding not about zero momentum but around points where some
components of $p$ are approximately $\pi$.  Thus, in addition to the
excitations around $p_\mu=0$, there are 15 other points in momentum
space where the action is small.  Visualizing momentum space as a
direct product of two toroids, the zeros occur as sketched in
Fig.~(\ref{fig1}).

\section{A unitary transformation}

\begin{figure*}
\centering
\includegraphics[width=3in]{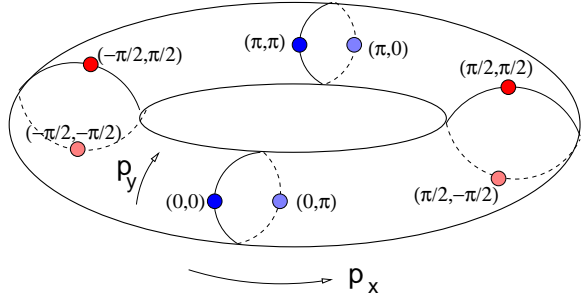}
\caption{The points in momentum space furthest from those giving the
  Dirac equation are located at $p_\mu=\pm\pi/2$.  Our second naive
  fermion action will have its zeros at these points.}
\label{fig2} 
\end{figure*}

I now do a transformation on the above action to generate a
superficially different but physically equivalent naive fermion
action.  Begin by considering maximally distant momenta from the zeros
of $D$.  There are 16 such points, occurring whenever the components
of the momentum satisfy $p_\mu=\pm\pi/2$.  These points are sketched
in Fig.~(\ref{fig2}) for the $x,y$ sub-torus.  Arbitrarily select one
of these points; here I consider $p_\mu=+\pi/2$ for all $\mu$.  Here
the original action becomes
\begin{equation}
D(p_\mu=\pi/2)= 2i \sum_\mu \gamma_\mu = 4i\Gamma
\end{equation} 
where I define the quantity
\begin{equation}
\Gamma\equiv {1\over 2}( \gamma_1+\gamma_2+\gamma_3+\gamma_4).
\end{equation} 
This is a unitary, Hermitean, and traceless four by four matrix in
spinor space.

Now consider a unitary transformation on the original fields
\begin{eqnarray}
&\psi^\prime= e^{-i\pi (x_1+x_2+x_3+x_4)/2}\ \Gamma\ \psi\\
&\overline\psi^\prime=
e^{i\pi(x_1+x_2+x_3+x_4)/2}
\ \overline\psi\ \Gamma. 
\end{eqnarray}
Here the $x_\mu$ are the integer coordinates of the lattice.  The
phases move the zeros in momentum space from $p_\mu=0,\pi$ to the
maximally distant points $p_\mu=\pm\pi/2$.  The factors of $\Gamma$
modify the gamma matrices for the new action to
\begin{equation}
\gamma_\mu^\prime=\Gamma\gamma_\mu \Gamma.
\end{equation}
Note that I can construct $\Gamma$ either from the original
$\gamma_\mu$ or the new $\gamma_\mu^\prime$
\begin{equation}
\Gamma={1\over 2}( \gamma_1+\gamma_2+\gamma_3+\gamma_4)
={1\over 2}
( \gamma_1^\prime+\gamma_2^\prime+\gamma_3^\prime+\gamma_4^\prime)
=\Gamma^\prime.
\end{equation}

In the free field limit, momentum space again diagonalizes the new
action, call it $\overline D$,
\begin{equation}
\overline D(p)=2i \sum_\mu \gamma_\mu^\prime \sin(\pi/2-p_\mu).
\end{equation}
Of course since all I have done is a unitary transformation, the Dirac
operators $D$ and $\overline D$ are physically equivalent.  They also
satisfy a complementarity manifested in
\begin{equation}
D(p_\mu=\pi/2)=\overline D(p_\mu=0)=4i\Gamma.
\end{equation}
This property is crucial to the following construction.

\section{The minimally doubled action}

\begin{figure*}
\centering
\includegraphics[width=3in]{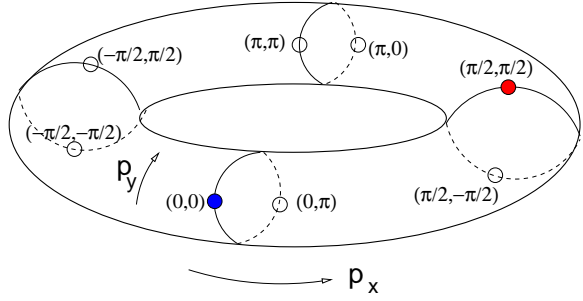}
\caption{The combination of terms ${\cal D} = D+\overline D-4i\Gamma$
leaves only two Fermi points in momentum space.  This is the minimal
number consistent with the Nielsen-Ninomiya theorem
\cite{Nielsen:1980rz}.}
\label{fig3} 
\end{figure*}

I now construct the final action from a linear combination of these
two equivalent theories
\begin{equation}
{\cal D} = D+\overline D-4i\Gamma.
\label{action}
\end{equation} 
To see how this works, consider the free theory limit in momentum
space
\begin{equation}
{\cal D}(p) =2i \sum_\mu\left(\gamma_\mu\sin(p_\mu)+\gamma_\mu^\prime
\sin(\pi/2-p_\mu)\right)-4i\Gamma.
\end{equation} 
Going to the Fermi point of the original theory at $p_\mu\sim 0$, then
the $4i\Gamma$ term cancels $\overline D$.  On the other hand, at the
Fermi point of $\overline D$ occurring at $p_\mu\sim \pi/2$, the
$4i\Gamma$ term cancels $D$.  The remarkable feature of this
combination is that only these two zeros of ${\cal D}(p)$ remain.  I
give a more detailed proof in Appendix A, but basically at the other
zeros of $D$, $\overline D-4i\Gamma$ is large and at the other zeros
of $\overline D$, $D-4i\Gamma$ is large.  Fig.~(\ref{fig3}) shows the
two remaining Fermi points.

Note that each term in Eq.~(\ref{action}) anti-commutes with
$\gamma_5$, maintaining the exact chiral symmetry of the naive action.
Thus a finite chiral rotation gives 
\begin{equation}
e^{i\theta\gamma_5}{\cal D} e^{i\theta\gamma_5}={\cal D}.
\label{rotation}
\end{equation}
The construction uses different gamma matrices for the two species.  A
particular consequence is $\gamma_5^\prime=\Gamma\gamma_5
\Gamma=-\gamma_5$.  Under the chiral rotation of Eq.~(\ref{rotation}),
the two species behave oppositely.  The symmetry is of a non-singlet
nature, as expected.

\section{Space time symmetries}

The above construction utilizes the matrix $\Gamma={1\over 2}\sum_\mu
\gamma_\mu$. Had I selected another of the maximally distant points
from the zeros of the initial fermion operator, then this relation
would be modified with minus signs for some directions.  This would
give an equivalent theory, but any such choice still involves picking
a particular diagonal axis of the fundamental hyper-cubic lattice as
special.  Here I have chosen the positive major diagonal.  For
comparison, in Refs.~\cite{Karsten:1981gd,Wilczek:1987kw}
the special direction is not
a diagonal but rather $x_4$, as discussed further in Appendix B.

Having picked a specific diagonal of the fundamental hypercubes, the
action is not symmetric under the full hyper-cubic group, but only the
subgroup that leaves this direction invariant.  This subgroup includes
$Z_3$ rotations that cycle any three positive axes.  For example, I
might want to cyclicly permute the $x_1,x_2,x_3$ axes.  To get the
gamma matrices to transform appropriately under such rotations,
introduce the matrix (in spinor space)
\begin{equation}
V=\exp\left({i\pi\over 3\sqrt 3} (\sigma_{12}+\sigma_{23}+\sigma_{31})\right).
\end{equation}
Here I define $[\gamma_\mu,\gamma_\nu]_+=2i\sigma_{\mu\nu}$.  The
combination $V^\dagger \gamma_\mu V$ cyclicly permutes the first three
gamma matrices.  Since $[V,\Gamma]=0$, this matrix, along with a
corresponding rotation of the gauge fields, realizes a symmetry of the
action.  Note that $V^3=-1$, indicating that a rotation by an angle of
$2\pi$ gives a minus sign, as expected for fermions.  Combining such
rotations using other axes generates the 12 element tetrahedral
subgroup of the hyper-cubic group.

The above rotations generate positive parity permutations of the
axes.  Introducing negative parity permutations increases the subgroup
to 24 elements.  For example, the matrix
\begin{equation}
V={1\over 2\sqrt 2}
(1+i\sigma_{15})(1+i\sigma_{21})(1+i\sigma_{52})
\end{equation} 
generates the fermionic part of a rotation that exchanges the $x_1$
and $x_2$ axes.  This transformation flips the sign of $\gamma_5$,
i.e.  $V^\dagger \gamma_5 V = -\gamma_5$, emphasizing that it is a
negative parity transformation.

In this formalism the natural direction to represent time is the main
diagonal $e_1+e_2+e_3+e_4$.  The combination of time reversal and
parity can then be chosen to flip the sign of all axes.  A unitarity
transformation similar to that relating $D$ and $\overline D$ restores
the action to its original form, increasing the symmetry group to 48
elements.

Note that charge conjugation symmetry is trivial in this formulation,
just being particle hole symmetry.  Both the operators $\cal D$ and
the Hermitean combination $\gamma_5 \cal D$ have their eigenvalues in
opposite sign pairs.

Because of the special treatment of the main diagonal, the effects of
interactions can distort lengths along this direction.  At each of the
Fermi points, this distortion is associated with the dimension five
continuum operator $i\overline \psi \Gamma\nabla^2\psi$.  Interactions
at finite lattice spacing can result in the gluons and fermions not
having the same speed of light.  This can be corrected with a
renormalization of the $\overline\psi \Gamma\psi$ term in the action,
as emphasized in Ref.~\cite{maryland}.  Nevertheless, the zeros are
stable under this distortion because they involve a topologically
non-trivial mapping of surfaces of constant action
\cite{Nielsen:1980rz,Creutz:2007af}.

\section*{Appendix A: Proof that there are only two zeros of ${\cal D}(p)$}

From the definition of $\Gamma$ it is elementary to show that
$\gamma^\prime_\mu=\Gamma-\gamma_\mu$ and the properties ${\rm
Tr}\Gamma \gamma_\mu={\rm Tr}\Gamma \gamma_\mu^\prime=2.$ Using
trivial trigonometric identities one can deduce
\begin{equation}
{\rm Tr}\ (\gamma_\mu-\gamma_\nu){\cal D}(p)
\sim \sin(p_\mu-\pi/4)-\sin(p_\nu-\pi/4).
\end{equation}
This implies that at any zero $\cos(p_\mu-\pi/4)=\pm\cos(p_\nu-\pi/4)$.
But a zero also requires
\begin{equation}
{\rm Tr}\ \Gamma{\cal D}(p)=0\ \Rightarrow
\ \sum_\mu\cos(p_\mu-\pi/4)=2\sqrt 2>2.
\end{equation}
Thus all these cosines must be positive.  Therefore all components of $p_\mu$
are equal and either 0 or $\pi/2$, the two Fermi points of interest.

\section*{Appendix B: Comparison with actions from Karsten and
  Wilczek}

Two other minimally doubled chiral fermion actions have been presented
in Refs.~\cite{Karsten:1981gd,Wilczek:1987kw}.  These actions are in
fact equivalent to each other under a unitarity transformation
$\psi\rightarrow e^{-i\pi x_4/2}\psi$.  For the free case, that action
can be written
\begin{equation}
D=\sum_{\mu=1}^4 \gamma_\mu \sin(p_\mu) + \gamma_4 \sum_{i=1}^3 (1-\cos(p_i)).
\end{equation}
The last term removes all doublers from the naive action except those
at  $\vec p=0$ and $p_4=0,\pi$.

The main difference from the action presented here is that $x_4$ is
now chosen as the special direction.  The on site term is proportional
to $\gamma_4$ instead of $\Gamma$.  As before the second species with
momentum around $p_4=\pi$ uses different effective gamma matrices,
${\vec \gamma}^\prime=\vec \gamma$ and $\gamma_4^\prime=-\gamma_4$.
As with the action from the main text, this gives
$\gamma_5^\prime=-\gamma_5$.  Again the chiral symmetry is flavored.
The symmetry of this system is the subgroup of the hyper-cubic group
that preserves the fourth axis up to a sign.

\end{document}